\documentclass[useAMS,usenatbib]{mn2e}
\usepackage{psfig}
\newcommand{\rareco}{${\rm C^{17}O}~J=2\rightarrow 1$}
\newcommand{\hco}{${\rm HCO^+}~J=3\rightarrow 2$}

\begin{document}
\title[Rotation of L1689B]
{Rotation of the pre-stellar core L1689B}
\author[Redman et al]
{M.P. Redman$^{1,2}$, E. Keto$^3$, J.M.C. Rawlings$^1$,
D.A. Williams$^1$\\ $^1$Department~of~Physics~and~Astronomy,
University~College London, Gower Street,London WC1E 6BT, UK\\ $^2$
School of Cosmic Physics, Dublin Institute for Advanced Studies, 5
Merrion Square, Dublin 2, Republic of Ireland\\
$^3$Harvard-Smithsonian Center for Astrophysics, 60 Garden Street,
Cambridge MA 02138, USA }
\date{\today}
\pubyear{2004} 
\volume{000}
\pagerange{\pageref{firstpage}--\pageref{lastpage}}
\maketitle 
\label{firstpage}

\begin{abstract}
The search for the onset of star formation in pre-stellar cores has
focussed on the identification of an infall signature in the molecular
line profiles of tracer species. The classic infall signature is a
double peaked line profile with an asymmetry in the strength of the
peaks such that the blue peak is stronger. L1689B is a pre-stellar
core and infall candidate but new JCMT HCO$^+$ line profile data,
presented here, confirms that both blue and red asymmetric line
profiles are present in this source. Moreover, a dividing line can be
drawn between the locations where each type of profile is found. It is
argued that it is unlikely that the line profiles can be interpreted
with simple models of infall or outflow and that rotation of the inner
regions is the most likely explanation. A rotational model is
developed in detail with a new 3D molecular line transport code and it
is found that the best type of model is one in which the rotational
velocity profile is in between solid body and Keplerian. It is firstly
shown that red and blue asymmetric line profiles can be generated with
a rotation model entirely in the absence of any infall motion. The
model is then quantitively compared with the JCMT data and an
iteration over a range of parameters is performed to minmize the
difference between the data and model. The results indicate that
rotation can dominate the line profile shape even before the onset of
infall.
\end{abstract}

\begin{keywords}
radiative transfer - submillimetre - stars: formation - ISM: clouds -
ISM: molecules - ISM: individual: L1689B
\end{keywords}

\section{Introduction}
The physics of the collapse process that leads to the formation of a
star is still not understood. Although many models of star formation
have been proposed, observational difficulties have limited the degree
to which they can be tested. Molecular line profiles, emitted in the
mm and sub-mm regimes, potentially offer the best opportunity to
extract dynamical information about the collapse process that leads to
the formation of stars. For some species the abundances are great
enough that self-absorption of the line takes place and if the gas is
undergoing infall then the line can assume a characteristic
double-peaked profile with a stronger blue wing. \citet{evans99} has
reviewed how such line profiles are generated and this is discussed
further below. Much observational effort has been concentrated on
detecting and characterising blue asymmetric line profiles in the hope
of using them to extract dynamical information. For example,
\citet{choi.et.al95} and \citet{zhou.et.al93} modelled the pre-stellar
core B335 as undergoing a \citet{shu77} 'inside-out'
collapse. However, the observational data in these cases consisted of
one or a few molecular lines at a single position in the cloud. Though
a consistent fit to the data could be made using a Shu collapse model
there was no guarantee that the fit was unique and
\citet{wilner.et.al00} showed that such a model was not able to
explain their ${\rm CS}~(5-4)$ data of B335. While the search for
infall in pre-stellar cores has become focussed on identifying blue
asymmetric line profiles, it is becoming clear that the interpretation
of the molecular line profiles from collapse candidates is much more
complex than previously thought with the shape of the line profile
being highly sensitive to the chemical and physical properties of the
gas (e.g. \citealt{rawlings&yates01,ward-thompson&buckley01}).

In this paper, an additional possible cause of asymmetric line
profiles is considered: rotation, which is suggested by the observed
spatial separation of red and blue asymmetric profiles in JCMT line
profiles obtained from the pre-stellar core L1689B, reported
below. L1689B is modelled as a system in which the central regions are
dynamically active and undergoing rotation while the outer regions are
quiescent. A 3D molecular line transport code is employed to model the
core and to directly compare the results with recently obtained JCMT
data. In Section 2 observations of L1689B are reviewed and the JCMT
data that is to be modelled is presented. The model and code are
described in Section 3. The results are presented and discussed in
Section 4 and compared with other recent studies, particularly that of
\citet{belloche.et.al02} who have recently demonstrated the presence
of rotation in the more evolved Class 0 source, IRAM 04191+1522. It is
concluded in Section 5 that the asymmetric line profiles can be very
well modelled if L1689B is undergoing rotation and that therefore
rotation dominates any infall motions in this source.

\section{The pre-stellar core L1689B}
L1689B is a pre-stellar core located in
Ophiuchus. \citet{shirley.et.al00} present SCUBA maps of L1689B that
confirm that there is no protostellar object in the centre of the
core. \citet{evans.et.al01} use a self-consistent dust emission model
to constrain the properties of L1689B and conclude that the dust
temperature decreases from the edge to the centre, a conclusion also
reached by \citet{andre.et.al96} and
\citet{jessop&wardthompson01}. The density structure of the core is
mildly flattened in the centre and decreases for larger radii. The
density was well modelled by \citet{evans.et.al01} using a
Bonner-Ebert sphere density distribution (see
e.\@g.\@~\citealt{mclaughlin&pudritz96};
\citealt{alves.et.al01};
\citealt{evans.et.al01}; \citealt{whitworth&wardthompson01};
\citealt{wuchterl&tscharnuter03} for recent work involving Bonner-Ebert 
spheres).

\citet{redman.et.al02b} 
carried out \rareco\ observations towards the pre-stellar core
L1689B. By examining the relative strengths of the hyperfine
components of this line the optical depth was calculated. This allowed
accurate CO column densities to be determined. The hydrogen column
densities that these measurements imply are substantially smaller than
those calculated from SCUBA dust emission data of
\citet{shirley.et.al00} and \citet{evans.et.al01}. Furthermore, the
\rareco\ column densities are approximately constant across L1689B
whereas the SCUBA column densities are peaked towards the centre. The
most likely explanation is that CO is depleted from the central
regions of L1689B, as also suggested by
\citet{jessop&wardthompson01}. Evidence of CO depletion has also been
found in several other prestellar cores
\citep{caselli.et.al99,bacmann.et.al02,jorgensen.et.al02,tafalla.et.al02}. It
was estimated by \citet{redman.et.al02b} that within about 5000 AU of
the centre of L1689B, over 90\% of the CO has frozen onto grains.
This level of depletion can only be achieved after a duration of time
that is at least comparable to the free-fall timescale.  While the CO
emitting gas appears to be quiescent, line profiles obtained by
\citet{gregersen&evans00} show the presence of both blue {\it and} red
asymmetric line profiles in this source. New JCMT observations,
described below, confirm the observational results of
\citet{gregersen&evans00} and reveal an underlying order to the
locations of the blue and red asymmetric profiles.

\section{JCMT observations of L1689B}
The bulk of the observations were carried out using the heterodyne
receiver RxA3 at the James Clark Maxwell Telescope (JCMT), Mauna Kea,
Hawaii on the nights of 2001 Feburary 24-26, 2001 August 19-21, 2003
May 3-5. Additional data were collected on various dates in
2001. \hco\ (267.5665 GHZ) line profiles were obtained for seventeen
positions across the centre of L1689B.  Some of the data points were
collected as part of a larger survey of infall candidate objects, the
results of which will be reported in Redman et al. (2004). The data
were reduced in the standard manner using the {\sc specx} software
package and were converted to the $T_{\rm mb}$ scale using a main beam
efficiency of 0.7.

\begin{figure*}
\psfig{file=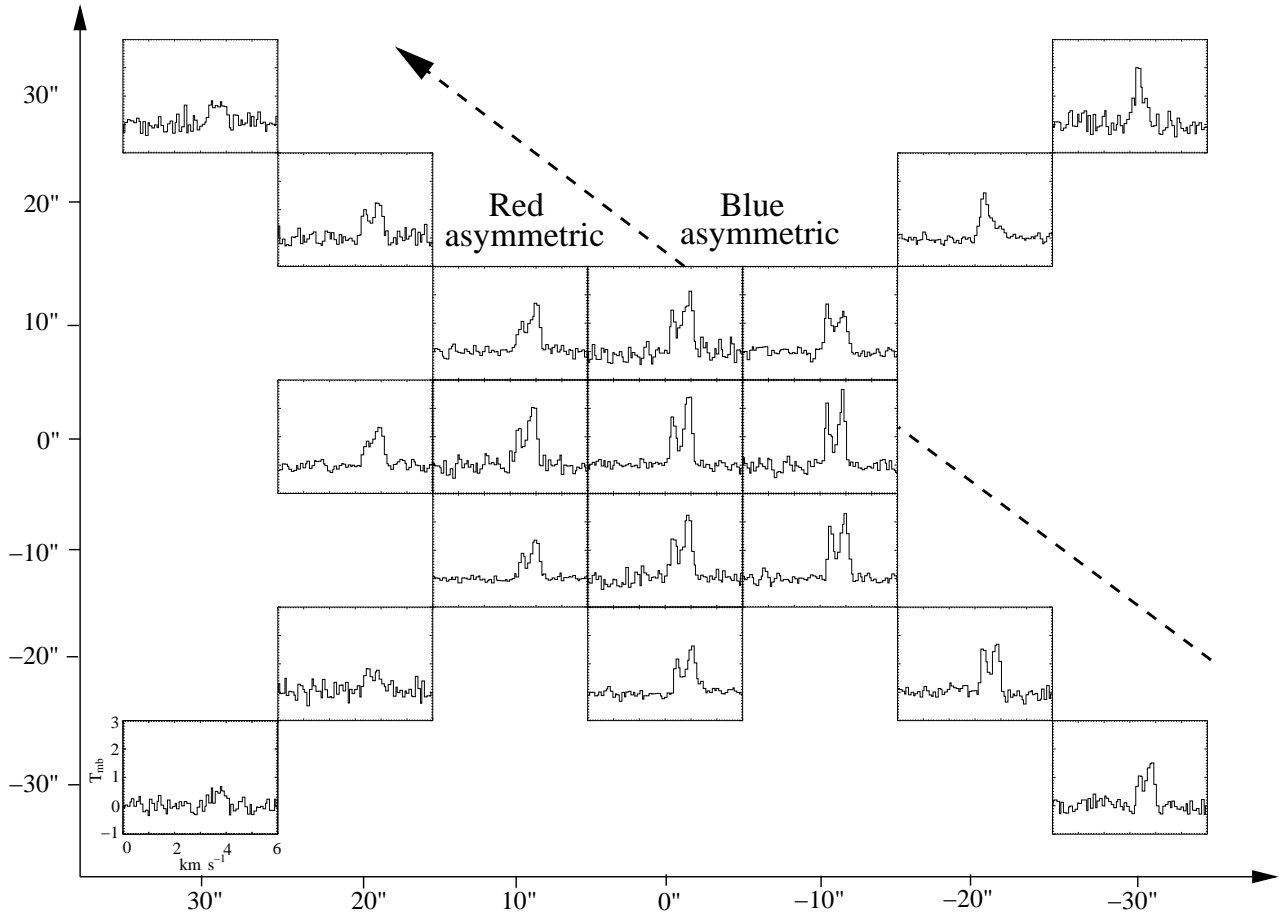,width=475pt,angle=0,bbllx=0pt,bblly=0pt,bburx=795pt,bbury=575pt}
\caption{\hco\ line profiles from positions across the centre of L1689B. The coordinates of
the pointing centre are RA 16 31 47.0, DEC -24 31 45.0 (B1950). The axis dividing blue and red asymmetric profiles is displaced to the NW of the zero offset position of our dataset}
\label{data}
\end{figure*}

The seventeen positions are at different offsets and at shorter
spacings than those of \citet{gregersen&evans00}. The line profiles
are displayed in Fig~\ref{data}.  Again, the presence of both blue and
red asymmetric line profiles is apparent. The two datasets cover
different, though overlapping, parts of the cloud (note that, 
as is often the case for prestellar cores, the position of the centre of 
the cloud is somewhat uncertain - see, e.g., the SCUBA dust emission map of Shirley et al 2000). While in our map
the majority of the profiles are red asymmetric line profiles, the
opposite is the case for the \citet{gregersen&evans00} dataset. Where
the pointing positions between the
\citet{gregersen&evans00} dataset and the present study overlap (in
the north-east [top right] corner) it was verified that the datasets
match each other.

Inspection of the data of \cite{gregersen&evans00} and Fig~\ref{data}
reveals that the blue and red asymmetric line profiles are found on
either side of an axis that runs from SW to NE. This axis is displaced to 
the NW of the zero offset position of our dataset but would run through the centre 
of the combined datasets. The change in asymmetry is in broad accord with the 
lower spatial resolution ${\rm
CS~{2\rightarrow 1}}$ data of \citet{lee.et.al99} (obtained with
FCRAO) in which the sense of asymmetry of the line profiles switches
from blue in the north of L1689B to red in the south. This indicates
that there is dynamical activity in the ${\rm HCO^+}$ emitting
gas. Since the outer layers of the cloud, as traced by ${\rm
C^{17}O}$, are quiescent (the hyperfine structure is well resolved -
\citealt{redman.et.al02b}) the dynamically active gas must lie within
the CO depleted region.

\section{Modelling}
A blue asymmetric line profile in an infalling core will be formed if
(i) the infall velocity increases less slowly than $1/r$ and (ii) the
excitation temperature increases towards the centre
(e.g. \citealt{evans99}). The dust radiative transfer modelling
carried out by \citet{evans.et.al01} demonstrates that the second of
these conditions may not be met in an object like L1689B since the
dust temperature (and, if they are well coupled, the gas temperature)
decreases towards the centre of the cloud. Furthermore, of course, the
presence of red asymmetric line profiles as well as blue asymmetric
profiles toward the centre of the cloud means that simple infall
cannot explain the observational data. 

\begin{figure}
\psfig{file=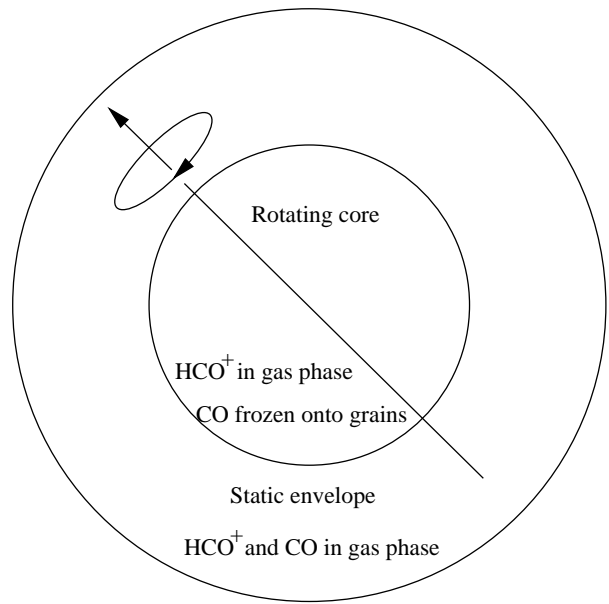,width=225pt,bbllx=0pt,bblly=0pt,bburx=408pt,bbury=408pt}
\caption{Sketch of the model of L1689B. The radius of the rotating centre is 3000 AU. \citet{redman.et.al02b} estimated that most of the CO is frozen onto grains within the central 5000 AU}
\label{sketch}
\end{figure}

Since star formation is usually accompanied by outflows, one possible
explanation is that an outflow is the cause of the changes in the
sense of the asymmetry. However, this can be discounted here since the
CO would be kinematically disturbed and the sharply resolved hyperfine
structure would be blended away. Furthermore, in a starless core with
an interior depleted in heavy molecules, such as L1689B, an outflow is
not yet expected to have been initiated. Instead, we model the system
as consisting of a quiescent outer halo where the CO is emitted, with
a rotating inner ${\rm HCO^+}$ emitting region. Figure~\ref{sketch} is
a sketch of the model.

\subsection{General description of the code}
In dark molecular clouds, the density is usually too low for local
thermodynamic equilibrium (LTE) to apply, the opacity is usually too high for
an optically thin approximation and the systemic velocities are too
low for the large velocity gradient (LVG) or Sobolev approximations to
be valid. We therefore generate model spectra and intensity maps
using a non-LTE numerical radiative transfer code (described in
\citealt{keto.et.al03} and also used by \citealt{rawlings.et.al04}) 
The code is fully 3 dimensional and employs the accelerated lambda
iteration (ALI) algorithm of \citet{rybicki&hummer91} to solve the
molecular line transport problem within an object of arbitrary
(three-dimensional) geometry.  This represents a considerable
improvement over the early models of
\citet{keto90}.

The local line profile is specified by systemic line-of-sight motions
together with thermal and turbulent broadening. In addition to
essential molecular data (such as collisional excitation rates) the
physical input to the model consists of the three-dimensional
velocity, density and temperature structures. The radiative transfer
equations are then numerically integrated over a grid of points
representing sky positions. The main source of uncertainty in the
radiative transfer calculations are the collisional rate coefficients,
but this should only be manifest in the uncertainties in the line
strengths (which may be inaccurate by as much as 25\%); the line
profile shapes and and relative strengths should be less affected.

For our purposes we define a spherical cloud within a regularly spaced
cartesian grid of $30\times 30\times 30$ cells. Within each cell, the
temperature, ${\rm H}_2$ density, molecular species abundance,
turbulent velocity width, and bulk velocity are specified. Externally,
the ambient radiation field is taken to be the cosmic microwave
background. At a given viewing angle to the grid, the line profile is
calculated at specified offsets from the centre by integrating the
emission along the lines of sight.  It is then possible to regrid the
data in the map plane and convolve with the (typically Gaussian) beam
pattern of the telescope, which for the JCMT is $15\arcsec$. The
spectra are computed with a frequency resolution of $0.01~{\rm
km~s^{-1}}$.

\subsection{Comparison with observational data}
Once a plausible general model has been set up, observational data can
be read directly into the code and multiple models with varying
parameters can be calculated and the difference between the model and
the data minimized. The dimension of the model cube is selected so
that the data points are positioned at integer values of the cube
spacing. The data is regridded so that the data channels match those
of the model output. After generating a model cube, the code compares
the model with the data and calculates $\chi^2$ for that fit. The
numerical techniques of simplex and simulated annealing are used to
drive the refining process where the code adopts new parameters and
calculates a new cube; full details are given in
\citet{keto.et.al03}. A typical fitting run may require the
calculation of several hundred model cubes.

\subsection{L1689B model}
The model parameters are listed in Table~\ref{parameters} along with
either the fixed adopted values used in the code or the range over
which a parameter was allowed to vary. The number of model runs
required to explore the parameter space increases rapidly with the
number of free parameters so those parameters which have the best
empirical contraints were given fixed values.
\begin{table}
\begin{tabular}{ll}
Parameter & Adopted value or range (best fit value)\\
\hline
Outer radius & $6000~{\rm AU}$\\
Plummer radius $r_0$ & 1000~{\rm AU}\\
Plummer density $\rho_0$ & $10^6~{\rm cm^{-3}}$\\
Temperature & $10~{\rm K}$ (edge) - $7~{\rm K}$ (centre)\\
$\rm HCO^+$ abundance & $5.0\times 10^{-10}-1.0\times 10^{-8}$ ($7.84\times 10^{-9}$)\\
Projected velocity & $0.1-0.4~{\rm km~s^{-1}}$ ($0.121$)\\
Rotation radius & $3000~{\rm AU}$ \\
Turbulent velocity & $0.12-0.18~{\rm km~s^{-1}}$ ($0.157$)\\
\end{tabular}
\label{parameters}
\caption{Model parameters used. For the parameters that were allowed to vary,
the best fit value is given in brackets}
\end{table}

The density profile used is that of a Plummer sphere which is defined
as
\begin{equation}
\rho(r)=\frac{\rho_0 r_0^2}{r^2+r_0^2}
\end{equation}
where $\rho_0$ is the central density and $r_0$ is the Plummer
radius. This simple density law has the property that as $r\rightarrow
0$, $\rho(r)\rightarrow \rho_0$ while for large $r$,
$\rho(r)\rightarrow {\rho_0 / r^2}$. The outer radius of the cloud is
defined as 6000 AU. The Plummer profile is also a reasonable
approximation to (and easier to deal with than) the Bonner-Ebert
sphere density distribution. This latter distribution appears to
describe the density profile of pre-stellar cores and is begining to
be used routinely in modelling work. The adopted values of $r_0$ and
$\rho_0$ were $1000~{\rm AU}$ and $10^6~{\rm cm^{-3}}$
respectively. These values were chosen to approximate the Bonner-Ebert
sphere fit to the density distribution of L1689B carried out by
\citet{evans.et.al01}. Similarly, an empirical approximation to the dust 
temperature profile fit of \citet{evans.et.al01} was employed to trace
the gas temperature (the gas density is sufficiently high that the two
temperatures can be considered well coupled). The temperature is 10 K
in most of the cloud and then drops, over a short distance, to 7 K in
the centre. In fact, trial runs with a constant temperature of 10 K
throughout the cloud yield very similar results to this temperature
profile.

The fractional abundance of ${\rm HCO^+}$ was taken to have a constant
value from the range $5.0\times 10^{-10}-1.0\times 10^{-8}$. The
assumption of a constant abundance is of course a crude approximation
and a self-consistent chemistry should be included in the
cloud. However, the chemistry of ${\rm HCO^+}$ is complicated and
\citet{rawlings.et.al92} show that its abundance intially rises as
depletion ensues. The sticking coefficient of ${\rm HCO^+}$ on to
charged and neutral dust grains is also not well known. Therefore an
attempt to include the chemistry does not yet seem worthwhile.

A simple solid body rotation law was adopted for the inner regions of
the model cloud. A rotation axis that matches the dividing line
between the red and blue asymmetric profiles was imposed. The range of
projected velocities investigated is $0.1-0.4~{\rm km~s^{-1}}$. The
rotation radius is an estimate of where the inner regions stop
rotating and is estimated as being at $3000~{\rm AU}$
(cf.\@~$5000~{\rm AU}$ for the CO depleted 'hole'
\citealt{redman.et.al02b}). Finally, the uncertain turbulent velocity
width was allowed to vary over the range $0.12-0.18~{\rm km~s^{-1}}$.

\section{Results and discussion}
\begin{figure*}
\psfig{file=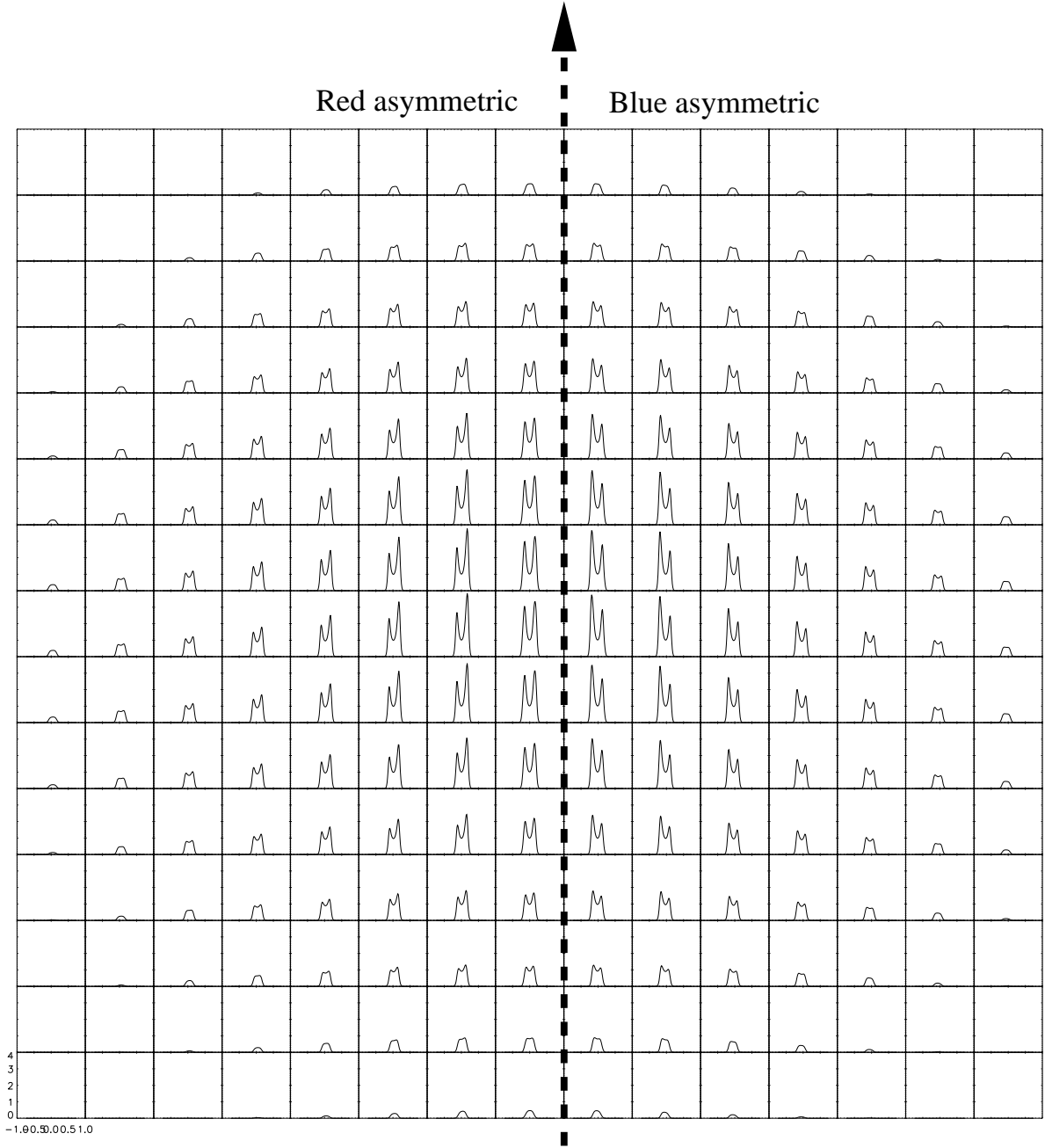,width=450pt,angle=0,bbllx=0pt,bblly=0pt,bburx=548pt,bbury=614pt}
\caption{${\rm HCO^+} J=3\rightarrow 2$ line profiles from the rotation model. For clarity the array of profiles is displayed with the rotation axis as shown. The line profiles switch between red asymmetric and blue asymmetric on either side of the rotation axis.}
\label{modelonly}
\end{figure*}

The full set of synthesised ${\rm HCO^+} J=3\rightarrow 2$ line
profiles for the best fit model is displayed in
Figure~\ref{modelonly}. For clarity, the rotation axis points
vertically. The line profiles show clear asymmetries with the sense of
the asymmetry switching between red and blue on either side of the
rotation axis. The self-absorption is strongest towards the centre of
the model cloud and the asymmetry is strongest there. Away from the
centre of the cloud, the drop in column density and self-absorption
results in a reduced line strength and less pronounced asymmetry (the
difference in strength between red and blue peak; the relative
magnitude of the absorption trough). These results confirm that it is
possible to generate both red and blue asymmetric line profiles
entirely in the absence of any infall or outflow motions.

\begin{figure*}
\psfig{file=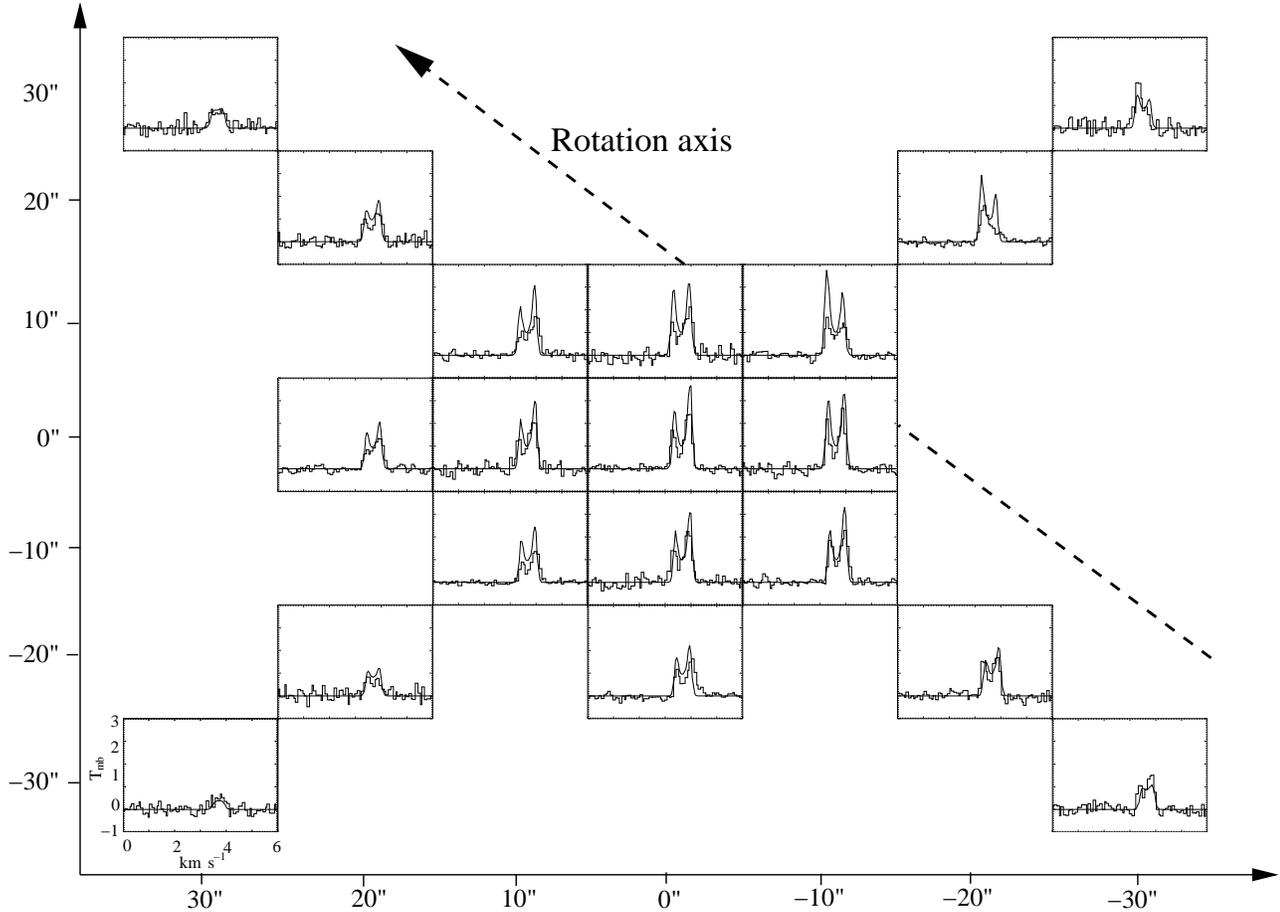,width=475pt,angle=0,bbllx=0pt,bblly=0pt,bburx=795pt,bbury=575pt}
\caption{${\rm HCO^+} J=3\rightarrow 2$ line profiles from L1689B. The spacing
between the cells is 10''. North is up and east is to the
right. Overlaid are synthesised line profiles generated by the
rotation model for the source.}
\label{hcocompare}
\end{figure*}

Figure~\ref{hcocompare} is an overlay of the best fit model described
above with the JCMT data of Redman et al (2004). The rotation axis is
marked on the figure. Given the simplicity of the model, the fit is
excellent with the sense of the asymmetry in the line profiles, their
shape and the fall-off in line strength away from the centre being
reproduced well. The most obvious of the discrepancies is that the
line strengths are overestimated for the brightest lines near the
centre of the cloud. One explanation for this may be due to the
freezing out of ${\rm HCO^+}$ onto grains.  Any freeze-out of the
${\rm HCO^+}$ is likely to occur first in the centre of the cloud,
where the density is highest, and progress outwards. If there is
significant freeze-out in the very centre of the cloud, this would
reduce the strength of the line profiles for small impact
parameters. As noted above, the chemical behaviour of ${\rm HCO^+}$ in
a core undergoing depletion is complicated.

\citet{belloche.et.al02} 
have carried out a detailed study of the Class 0 protostar IRAM
04191+1522 and conclude that the inner region of IRAM 04191+1522 is
undergoing fast differential rotation while the outer regions are only
slowly rotating.  Since they also argue that this source has only
recently become a protostar it is constructive to compare it with
L1689B which may be on the verge of collapse.  Firstly, under the
assumption that both objects have been adequately modelled, the
angular velocity is comparable: $\omega r=9.4~{\rm km~s^{-1}}$ at 3000
AU in the best fit model for L1689B while for IRAM 04191+1522, $\omega r=9\pm
3~{\rm km~s^{-1}}$ at 2800 AU. IRAM 04191+1522 may be decoupling from the
outer envelope at a radius of 2000-4000 AU \citep{belloche.et.al02}
which compares with the rotation radius of 3000 AU estimated here for
L1689B. Solid body rotation is firmly ruled out for the inner region
of IRAM 04191+1522 with differential rotation providing a much better fit
to the observations (specifically, the angular velocity gradient is
found to change from approximately $\omega\propto r^{-2.5}$ to
$\omega\propto r^{-1}$ in the inner regions with very slow rotation in
the outer regions). The model results presented here for L1689B also
supports an $\omega\propto r^{-1}$ angular velocity profile. Such a
velocity gradient indicates that angular momentum is conserved and
that magnetic braking does not dominate the dynamics;
\citet{belloche.et.al02} provide an extensive discussion of their
results in terms of various collapse models. Thus, L1689B exhibits
strong similarities with IRAM 04191+1522 in terms of the rotational
behaviour. The principal kinematic difference is that IRAM 04191+1522 shows
clear evidence of infall (and outflow) in addition to rotation yet the
signature of any such motion in L1689B is completely dominated by the
rotation. It seems reasonable to suggest that L1689B and IRAM 04191+1522
are similar objects that are at slightly different evolutionary
stages, either side of the onset of infall.

\section{Conclusions}
A model of the pre-stellar collapse candidate L1689B has been presented
where the inner regions of the cloud are undergoing rotation while the
outer part of the cloud is static and quiescent. The model
calculations demonstrate that blue and red asymmetric line profiles
arise naturally from such a system even in the absence of infall or
outflowing gas motions. The possibility that rotation is present in a
pre-stellar object must therefore be addressed if asymmetric line
profiles are to be used to argue that a core is undergoing infall.

A best fit model was generated by varying several of the free
parameters within reasonable ranges. An excellent fit to the data was
obtained and we would therefore argue that our model is a plausible
one for L1689B. 

This study along with other recent work demonstrates that great care
needs to be taken in interpreting asymmetric line profiles as evidence
for infall. While this means that earlier claims for detection of
infall (e.\@g.\@ \citealt{zhou.et.al90}) ought now be re-evaluated it
does not mean the search will be fruitless. A combination of chemical
models, radiative transfer codes and high quality observations should
soon lead to the first firm identification of infall and the long
awaited detailed examination of the competing star formation models.

\section*{Acknowledgements}
MPR was funded by PPARC while this work was carried out. DAW
acknowledges the support of the Leverhulme Trust. This work was
partially supported by the National Science Foundation through a grant
for the Institute for Theoretical Atomic, Molecular and Optical
Physics at Harvard University and Smithsonian Astrophysical
Observatory. We thank the staff of the JCMT for their excellent
assistance during the observations. The JCMT is operated by the JAC,
Hawaii, on behalf of the UK PPARC, the Netherlands NWO, and the
Canadian NRC.


\label{lastpage}
\end{document}